\documentclass[twocolumn]{revtex4-1} 
\usepackage{graphicx,amsmath,amssymb,mathrsfs,eufrak,dcolumn,bm}
\begin{document}
\title{Vacancy-Induced Topological Fano Resonance in Kane-Mele Nanoribbons: Design,
Control, and Sensing Applications}
\author{S. Jalilvand$^1$}
\author{M. Soltani$^1$}\email{mo.soltani@sci.ui.ac.ir}
\author{Z. Noorinejad$^2$}
\author{M. Amini$^1$}
\author{E. Ghanbari-Adivi$^1$}\email{ghanbari@phys.ui.ac.ir}
\affiliation{$1$ Faculty of Physics, University of Isfahan, Isfahan
81746-73441, Iran}
\affiliation{$2$ Department of Physics,  Islamic Azad University -
Shahreza Branch, Shahreza, Iran}
\begin{abstract}
The concept of topological Fano resonance, characterized by an
ultrasharp asymmetric line shape, is a promising candidate for
robust sensing applications due to its sensitivity to external
parameters and immunity to structural disorder. In this study, the
vacancy-induced topological Fano resonance in a nanoribbon made up
of a hexagonal lattice with armchair sides is examined by
introducing several on-site vacancies, which are deliberately
created at regular distances,  along a zigzag chain that stretches
across the width of the ribbon. The presence of the on-site
vacancies can create localized energy states within the electronic
band structure, leading to the formation of an impurity band, which
can result in Fano resonance phenomena by forming a conductivity
channel between the edge modes on both armchair sides of the ribbon.
Consequently, an ultracompact tunable on-chip integrated topological
Fano resonance derived from  the graphene-based nanomechanical
phononic crystals is proposed. The Fano resonance arises from the
interference between topologically protected even and odd edge modes
at the interface between trivial and nontrivial insulators in a
nanoribbon structure governed by the Kane-Mele model describing the
quantum spin Hall effect in hexagonal lattices. The simulation of
the topological Fano resonance is performed analytically using the
Lippmann-Schwinger scattering formulation. One of the advantages of
the present study is that the related calculations are carried out
analytically, and in addition to the simplicity and directness, it
reproduces the results obtained from the Landauer-B\"{u}ttiker
formulation very well both quantitatively and qualitatively.  The
findings open up possibilities for the design of highly sensitive
and accurate robust sensors for detecting extremely tiny forces,
masses, and spatial positions.
\end{abstract}
\maketitle
\section{Introduction~\label{Sec01}}
Fano resonance~\cite{Fano01} is a quantum phenomenon that occurs
when a discrete energy level interacts with a continuum of energy
states, resulting in the formation of a characteristic asymmetric
line shape known as the Fano profile in the transmission or
absorption spectrum. The distinctive shape arises from the
constructive and destructive interference between the discrete state
and the states in the continuum. Fano resonances have been observed
in various physical systems, including atomic and solid-state
physics, electromagnetism, electronics, Aharonov-Bohm
interferometers, and quantum
dots~\cite{Miroshnichenko01,Ujsaghy01,Johnson01,Kobayashi01,Kamenetskii01,Attaran01,Lv01,Gores01}.
The characteristic shape of Fano resonances makes them useful in a
range of applications, including sensors, optical devices, and the
study of quantum interference phenomena. Recently, researchers have
introduced the concept of topological Fano resonance, whose
ultra-sharp asymmetric line shape is guaranteed by design and
protected against geometrical disorder, while remaining sensitive to
external parameters. Topologically protected Fano resonances have
been observed in various systems, including acoustic, microwave,
optical, and plasmonic systems, and they open up exciting frontiers
for the generation of various reliable wave-based devices, including
low-loss perfect absorbers, ultrafast switches or modulators, and
highly accurate interferometers, by circumventing the performance
degradations caused by inadvertent fabrication
flaws~\cite{Bandopadhyay01,Stassi01,Overviga01,Lukyanchuk01}.\par
In a lattice or crystal, electrons exhibit wave-like behavior. When
a defect is introduced, such as an impurity or a  vacancy, it
creates a localized state within the crystal's band structure. This
localized state has the potential to interact with the continuum of
extended states in the crystal, leading to the formation of Fano
interference. In this regard, studies utilizing discrete lattice
models have demonstrated the presence of Fano resonance in the
transmission spectrum of discrete lattice models with impurities
attached to the side~\cite{Deo01,Miroshnichenko02,Chakrabarti01}.
For instance, the electronic transmission through a one-dimensional
tight-binding lattice with a quantum dot coupled to the side to
explore the Fano anti-resonances in the transport through quantum
dots is studied in Refs.~\cite{Torio01,Rodriguez01,Guevara01}.
Recently, the theory for Fano anti-resonances induced by coupling
between vacancy states and edge states of the zigzag phosphorene
nanoribbons is investigated analytically~\cite{Amini01,Amini02}.\par
In contrast to conventional Fano resonances, which often suffer from
sensitivity to disorder and imperfections, the advent of topological
systems~\cite{Hasan01} featuring protected states presents a
promising alternative which is called topological Fano
resonance~\cite{ZangenehNejad01}. The topological protection ensures
that the unique features of Fano resonance, such as asymmetric line
shapes and sensitivity to environmental changes, remain intact even
in the presence of fabrication imperfections. This robustness can
lead to the development of more reliable and efficient devices,
including sensors, modulators, and interferometers, where the
extreme sensitivity of Fano resonances can be harnessed without
being compromised by fabrication
challenges~\cite{Ji01,Wang01,Sun01}. The concept of topological Fano
resonance and its potential applications in the development of
robust and reliable devices is an active area of research, and the
search results provide valuable insights into the control and
influence of impurities on Fano resonances in discrete lattice
models.\par
In this paper, the possibility of creating and controlling Fano
resonances in a system that exhibits the quantum spin Hall (QSH)
effect is explored. The considered system is a two-dimensional
topological phase of matter characterized by the presence of robust
and exotic edge states that are immune to local perturbations and
disorder. In regard of the task,  a nanoribbon of honeycomb
structure is considered in which the spin-orbit coupling is
described by the Kane-Mele model. This model is the first
theoretical model to predict the QSH effect in graphene. It is shown
that introducing a line of vacancy defect across the width of the
ribbon induces Fano anti-resonances in the transmission of the edge
states, which are topologically protected. In fact, the presence of
the on-site vacancies leads to emergence of some zero-energy
localized states surrounding the vacant sites. Overlapping these
created zero-energy localized states and their coupling with the
edge modes of the armchair sides generate an impurity band in the
ribbon's band structure. As a result, a conduction channel is formed
between the topologically protected edge modes localized at the
edges of these armchair sides. The presence of this channel causes
the edge modes on both sides of the ribbon to be connected.
Subsequently, this connection creates the substrate for occurring
the backscattering processes which causes the observation of the
topological Fano resonance in the scattering region created across
the width of the ribbon. Since, the Fano resonance is a widespread
wave scattering phenomenon, the Lippmann–Schwinger(LS)  equation
in the formal scattering theory along with the concepts such as the
Green function and the transition operator, can be employed to
calculate the transmission and backscattering probabilities of the
electrons during their passing through the channel between the edge
wires. Analytically, the topological Fano resonance simulation
utilizes the LS scattering formulation. This study offers an
advantage by conducting related calculations analytically, resulting
in outcomes that closely align with those achieved through the
Landauer-B\"{u}ttiker~(LB) formulation, demonstrating both
quantitative and qualitative accuracy.\par
Consequently, this research aims to investigate the design, control
and sensing applications of topological Fano resonances in Kane-Mele
nano ribbons, thereby  contributing to the advancement of nano scale
sensing technologies. Our results reveal a novel way to manipulate
and control the edge state transport in topological insulators, and
to create reliable and tunable Fano-based devices.\par
The article is organized as follows. In section 2, the geometry of
the specified system, the Kane-Mele Hamiltonian describing the
electronic properties of the topological insulators, and the
topological Fano resonance profile are explained to give a vision of
the vacancy-induced topological Fano resonance. In section 3, first
the discrete energy levels associated to a line of the on-site
vacancy defects are explained and discussed. Then, employing the
formal scattering theory and using the LS~equation as well as the
Green, and transition operators, the electron transmission and
backscattering probabilities in the scattering region created across
the width of the ribbon are calculated. The obtained results are
provided and discussed. Finally, in the last section, a summary
along with the conclusion remarks are presented.
\section{Vacancy-induced topological Fano resonances \label{Sec02}}
In this section, we delve into the intricacies of vacancy-induced
topological Fano resonances, exploring the nuanced interplay between
the topological characteristics of the Kane-Mele
model~\cite{Kane01,Kane02} and the localized states induced by these
engineered defects. Through systematic investigation, we unravel the
distinct features and behavior exhibited by the Fano resonances in
the presence of such vacancies, shedding light on their potential
implications for quantum transport and edge state engineering in
topological materials. The section unfolds with an examination of
the theoretical underpinnings, followed by a detailed presentation
of numerical results and insightful discussions on the observed
phenomena.
\begin{figure}[t!]
\begin{center}
\includegraphics[scale=0.46]{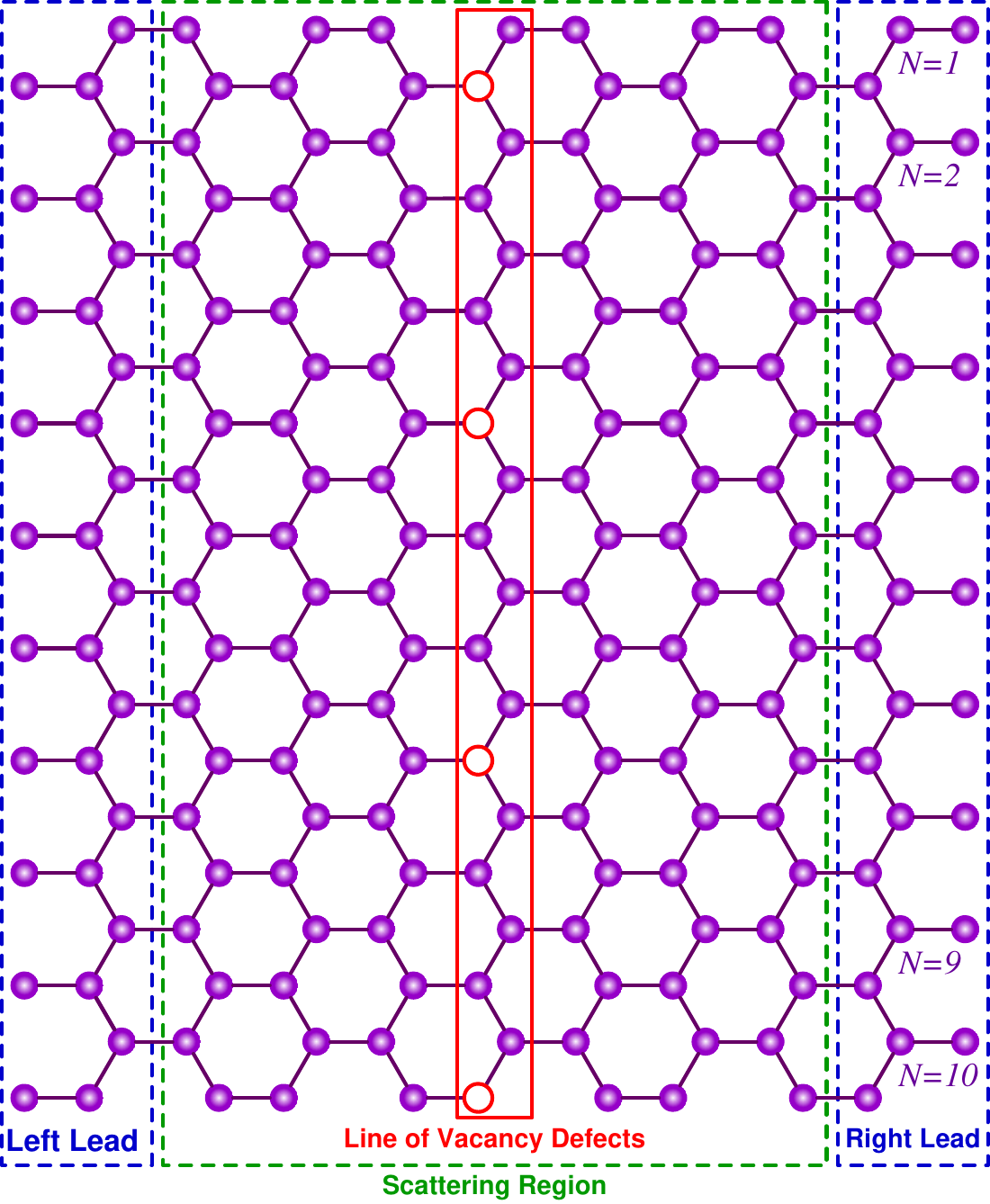}
\end{center}
\vspace{-5mm}\caption{Schematic representation of the Kane-Mele
nanoribbon with the honeycomb lattice structure. The ribbon is
infinite in the $x$-direction and finite in the $y$-direction  with
armchair edges. The system is connected to left and right leads. A
scattering region at the center hosts vacancy defects arranged on a
zigzag chain highlighted by a red rectangle box on the diagram.
Armchair chain numbers are labeled by $N$. The vacant sites are
denoted by open red circles which are introduced on armchair chains
of $N=1,\ 4,\ 7,\ 10$. \label{Fig01}}
\end{figure}
\subsection {Model Hamiltonian and system geometry\label{Sec02Sub01}}
The Kane-Mele model Hamiltonian captures the topological properties
of the system, considering the interplay between hopping terms,
spin-orbit coupling, and staggered on-site potentials. For a
honeycomb lattice ribbon with armchair edges, infinite in the
$x$-direction and finite in the $y$-direction, shown in
Fig.~\ref{Fig01}, the tight-binding Hamiltonian is given
by~\cite{Kane01,Kane02}:
\begin{equation}\label{Eq01}
\begin{split}
{\cal H}_{KM} & =  -t \sum_{\langle m, n \rangle, \sigma} c_{m,
\sigma}^{\dagger} c_{n, \sigma}\\ & - i \lambda_{SO} \sum_{\langle
\langle m, n \rangle \rangle, \sigma, \sigma'} e^{i\nu_{mn}\phi}
c_{m, \sigma}^{\dagger} (\boldsymbol{\sigma}_{\sigma, \sigma'})_z
c_{n, \sigma'}\\ & + M \sum_{m, \sigma} (-1)^m c_{m,
\sigma}^{\dagger} c_{m, \sigma}
\end{split}
\end{equation}
Here, $c_{m, \sigma}^\dagger~(c_{m, \sigma})$ creates~(annihilates)
an electron at site $i$ with spin $\sigma$, $t$ is the
nearest-neighbor hopping amplitude, $\lambda_{SO}$ is the spin-orbit
coupling strength, $M$ represents the staggered on-site potential,
and $e^{i\nu_{mn}\phi}$ is a phase factor in which $\nu_{mn} = \pm
1$ indicates the direction of the next-nearest-neighbor hopping. The
sums are taken over nearest neighbors $\langle m, n \rangle$ and
next-nearest neighbors $\langle \langle m, n \rangle \rangle$. ${\bf
\sigma} = \sigma_x{\bf e}_x +  \sigma_y{\bf e}_y$ is the well-known
Pauli spin operator.
The system is configured as a ribbon attached to left and right
leads, forming a scattering region in the center. Figure~\ref{Fig01}
illustrates the geometry with vacancy defects arranged in a line
which is shown by a red rectangle box. Importantly, placing a
vacancy defect on a given site is equivalent to turning off the
hopping terms that connect this site to its neighbors. These
vacancies play a pivotal role in inducing topological Fano
resonances, as will be explored in subsequent sections.
\subsection{The topological Fano profile~\label{Sec02Sub02}}
In the Kane-Mele model, the topological nature of the system
manifests in the existence of topologically protected edge states
within the energy gap. These edge states, forming helical edge
bands, give rise to an edge current that is impervious to
backscattering, a hallmark of the topological phase. In the absence
of mechanisms facilitating backscattering, the edge current remains
unaffected even in the presence of nonmagnetic defects like
vacancies. However, the introduction of vacancies in a specific
arrangement, forming a conduction channel that connects the upper
and lower edges of the ribbon, can lead to the emergence of
topological Fano resonance. This channel is effectively created by
engineering the defects in a line across the width of the ribbon. It
is noteworthy that the presence of vacancy defects, in contrast to
on-site impurities with finite potential~\cite{Rahmati01}, induces
discrete energy-bound states around zero energy.  Due to the
particle-hole symmetry of the system the energy  of these bound
states will be symmetrically distributed around $E=0$.\par
\begin{figure}[t!]
\begin{center}
\includegraphics[scale=0.72]{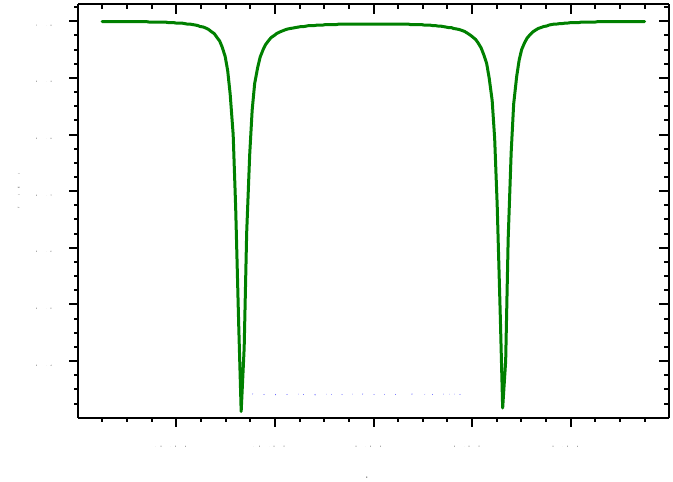}
\end{center}
\caption{Transmission spectrum through a Kane-Mele nanoribbon of
width $N = 10$, calculated using the LB formalism for conduction in
confined structures. The ribbon is in the topological phase with a
fixed hopping amplitude $t = 1$ as the energy unit, spin-orbit
coupling $\lambda_{\text{SO}} = 0.2$, and staggered potential $M =
0$. Four vacancy defects are strategically placed in a line on
armchair chains $m = 1, 4, 7, 10$, forming a conduction channel
connecting the upper and lower edges. The transmission spectrum
exhibits pairs of dips symmetrically distributed in the energy space
around $E = 0$, indicative of the topological Fano resonance.
\label{Fig02}}
\end{figure}

Our focus now shifts to the quantum transport through the ribbon in
the presence of vacancy defects. In current investigation, the
armchair chains are labeled by integer numbers $N=1~,2,~3,~\cdots$,
and four vacancy defects are introduced on a zigzag chain, situated
on armchair chains $N = 1,\ 4,\ 7,\ 10$ as illustrated in
Fig.~\ref{Fig01}. We specifically consider a ribbon with a width of
$N = 10$, indicating the number of armchair chains across the width.
The vacancies are regularly arranged so that the distance between
two consecutive vacant sites is equal to three zigzag chains.  In
order to check the accuracy of our analytical simulation of the
vacancy-induced topological Fano resenance in the specified
structure, which will be detailed in next section, we obtain the
transmission spectrum of the system by employing the LB formalism
for conduction in confined structures~\cite{Gamayun01}. For our
analysis, we set the hopping amplitude $t = 1$ as the unit of
energy, and to maintain the system in the topological phase, we
choose a fixed spin-orbit coupling of $\lambda_{\text{SO}} = 0.2$,
while keeping the staggered potential $M = 0$ for simplicity.  The
line of vacancy defects is highlighted by a red rectangle box. Our
findings reveal distinctive dips in the transmission spectrum
through the ribbon which is shown in Fig.~\ref{Fig02}. Intriguingly,
these dips appear in pairs and exhibit symmetrical distribution in
the energy space around $E = 0$. It is important to note that upon
closer examination of the profile, it is evident that the dips do
not exhibit symmetrical behavior with respect to the energies at
which they occur. Fano resonance phenomenon often reveals itself
through the asymmetrical dips in occurrence energies, a unique
fingerprint of this effect. This phenomenon signifies the occurrence
of topological Fano resonance in the Kane-Mele model, forming the
crux of our study, which we delve into further details in the
subsequent sections.\par
\section{Transmission analysis~\label{Sec03}}
In the presence of multiple vacancy defects arranged in a line, each
defect introduces a discrete energy level within the gap region of
the Kane-Mele nanoribbon. These localized states, associated with
the vacancies, exhibit localization in real space, forming a spatial
pattern that aligns with the arrangement of the defects. Crucially,
when defects are strategically placed, the associated states
overlap, creating a conduction channel that connects the upper and
lower edges of the ribbon. To understand the coupling of these
localized states with the edge states and the emergence of the
topological Fano resonance, we employ the LS~approach. This
theoretical framework allows us to obtain an explicit analytical
expression for the scattering amplitude of the edge electrons
through the presence of multiple vacancy defects. The resulting
scattering amplitude reveals a Fano resonance profile in the
transmission spectrum, characterized by tunable dips symmetrically
distributed around $E = 0$. The position and width of these
resonances are influenced by the arrangement of the vacancy defects,
offering a means to control and engineer the topological Fano
resonance in the Kane-Mele nanoribbon.\par
The combination of the discrete energy levels induced by vacancy
defects, their coupling to the edge bands, and the resulting Fano
resonance in the transmission spectrum forms the basis for our
exploration of vacancy-induced topological Fano resonances in the
Kane-Mele nanoribbon. In the subsequent sections, we will provide a
detailed analysis of these phenomena, shedding light on the rich
interplay between topology, defects, and quantum transport in
two-dimensional materials.\par
\subsection{Discrete energy levels associated with a line of vacancy defects~\label{Sec03Sub01}}
While the theorem governing the creation of zero energy modes in
bipartite lattices due to a single vacancy site is well-established
for honeycomb lattices in the absence of spin-orbit coupling
$(\lambda_{SO} = 0)$~\cite{Brouwer01,Pereira01}, its applicability
to more complex lattice structures, such as the Kane-Mele model,
requires further examination.\par
For graphene, where only nearest-neighbor hopping is considered, the
Hamiltonian can be represented in the matrix block form as
\begin{equation}\label{Eq02}
\Big(\begin{array}{cc}
{\bf 1} & H_{AB} \\
H_{AB}^\dagger & {\bf 1}
\end{array}\Big),
\end{equation}
satisfying the conditions for the creation of a localized zero
energy mode~\cite{Pereira01}. Here $A$ and $B$ refers to two
sublattices of the bipartite lattice, and $H_{AB}$ and  ${\bf 1}$
stand the hopping and identity matrices, respectively. The extension
of this result to materials like phosphorene, with next-nearest
hopping amplitudes, yields localized states with non-zero
energy~\cite{Amini01}. In our specific case of the Kane-Mele
Hamiltonian, ${\cal H}_{KM}$, the direct applicability of
Eq.~\eqref{Eq02} is not immediately evident. However, our numerical
simulations demonstrate that even in the presence of spin-orbit
coupling ($\lambda_{SO} \neq 0$), a single vacancy defect gives rise
to a zero-energy localized
state~\cite{Rahmati01,Sadeghizadeh01,AlShuwaili01}.\par
\begin{figure}[t!]
\begin{center}
\includegraphics[scale=0.42]{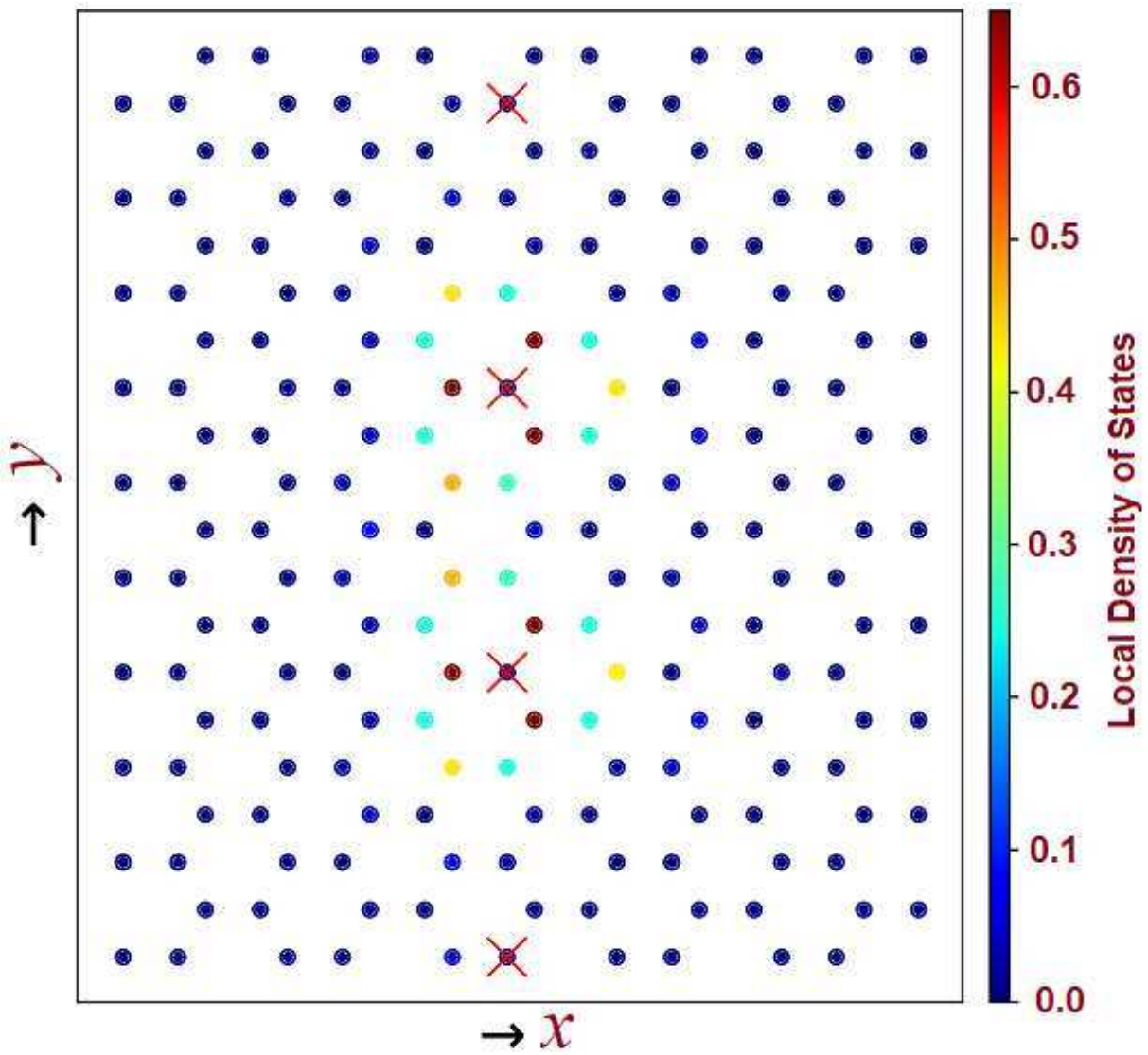}
\end{center}
\caption{The spatial profile of the local density of states~(LDOS)
corresponds to the configuration shown in Fig.~\ref{Fig01}, as it is
clearly visible from the figure, the presence of the vacancies
creates a conduction channel between the edge states at the upper
and lower armchair edges of the ribbon. This channel provides the
necessary platform for the occurrence of the backscattering events
and the observation of topological Fano resonance in the
transmission spectrum of the system. The vacancies  are marked with
cross symbols.~\label{Fig03}}
\end{figure}
To validate the mentioned behavior, the Kane-Mele model for the
honeycomb lattice was considered and the local density of states
around a single vacancy defect was examined. The obtained results
verify the emergence of a localized state with zero energy,
affirming the presence of vacancy-induced discrete energy
levels~\cite{Rahmati01,Sadeghizadeh01,AlShuwaili01}.\par
It is worth noting that the expressions for the edge states in the
presence of spin-orbit interaction~\cite{Rahmati01,Sadeghizadeh01}
become notably more intricate compared to the case without it
($\lambda_{\text{SO}} = 0$)~\cite{AlShuwaili01}. Consequently,
deriving an exact analytic expression for the localized wave
function associated with a single vacancy defect in the presence of
spin-orbit coupling ($\lambda_{\text{SO}} \neq 0$) would entail
increased complexity, surpassing the scope of this current study.
The detailed analysis of such wave functions remains an avenue for
future investigations.\par
Let us now proceed to the scenario of an infinite number of vacancy
defects periodically placed in the bulk honeycomb lattice. The
resultant band structure of the Kane-Mele model in the presence of
spin-orbit interaction has been examined in details by using both
the numerical and semi-analytical methods in
Refs.~\cite{Rahmati01,Sadeghizadeh01,AlShuwaili01}. Using the
information reported in these research works, it is possible to
examine the periodic arrangement involves placing the vacancy
defects such that there are three armchair chains between adjacent
defects, essentially mirroring the configuration illustrated in
Fig.~\ref{Fig01} but extending infinitely in the $y$-direction. For
this examination, a non-zero spin-orbit coupling strength of
($\lambda_{SO} \neq 0$) has been considered. The resulting band
structure reveals an additional energy band within the original
energy gap, attributed to the formation of an impurity band. Each
vacancy defect induces a localized wave function akin to a single
atomic orbital. These localized wave functions, resembling atomic
orbitals, overlap with neighboring wave functions from adjacent
vacancy defects, giving rise to the impurity band. Remarkably, the
obtained impurity band can be well-described by a simple
tight-binding band dispersion, represented as $E(k) = 2 \Delta
\sin(ka)$, where $\Delta$ and $a$ are the band width parameter and
the lattice constant, respectively.\par
The presence of a vacancy defect introduces a discrete energy level
at zero with a localized wave function reminiscent of a single-level
atom. In the case of a finite number of closely situated vacancy
defects, effective overlap occurs between these wave functions,
facilitating effective tunneling, $\Delta$, and resulting in finite
discrete energy levels symmetrically distributed around zero. These
energy levels exhibit extended wave functions across the width of
the ribbon, connecting the upper and lower edges and allowing for
backscattering. This phenomenon manifests as Fano anti-resoances in
the transmission spectrum of the system which will be described in
the following.\par
For a closer examination of the conductive channel created between
the edge modes at the upper and lower weirs around the ribbon
depicted in Fig.~\ref{Fig01}, the spatial profile of the local
density of states is displayed in Fig.~\ref{Fig03}. The illustrated
profile corresponds to that arrangement of the vacant sites shown in
Fig.~\ref{Fig01}. As can be seen from the figure evidently, the
overlap of the zero-energy localized modes around the vacant sites
forms a conduction channel between the upper and lower edges of the
ribbon. Consequently, the backscattering of the electrons during
their passing through this conductive channel, causes the
observation of the topological Fano resonance in the considered
system.
\subsection{Calculation of the Fano anti-resonances using the  Lippmann–Schwinger approach~\label{Sec03Sub02}}
The purpose of this section is to simulate the topological Fano
resonance observed in the transmission spectrum shown in
figure~\ref{Fig02}. To model the quantum transport through the
specified system and calculate Fano anti-resonances, the Hamiltonian
of
\begin{equation}\label{Eq03}
{\cal H} = {\cal H}_E + {\cal H}_V + {\cal H}_{EV},
\end{equation}
is employed in which Hamiltonian ${\cal H}_E$ corresponds to the
helical edge states of the Kane-Mele model and is expressed as:
\begin{equation}\label{Eq04}\begin{split}
{\cal H}_E & = \int_{-\pi}^{+\pi} \hbar {\rm v}_F k_U |k_U
\rangle\langle k_U| dk_U\\ &\qquad\qquad + \int_{-\pi}^{+\pi} \hbar
{\rm v}_F k_L |k_L \rangle\langle k_L| dk_L.\end{split}
\end{equation}
in which it is assumed that at the upper side, the continuum edge
state travels to the right with the Fermi speed of ${\rm v}_F$,
while at the lower side, it is traveling to the left at the same
speed. Additionally, the lattice constant, $a$, is assumed to be
one. Here, $k_U$ and $k_L$ represent, respectively, the wave vectors
for edge modes at the upper and lower edges associated to the ribbon
displayed in Fig.~\ref{Fig01}. Hereafter, the $U$ and $L$ indices
refer to the upper and lower edges.\par
The Hamiltonian ${\cal H}_V$ describes the line of vacancy defects:
\begin{equation}\label{Eq05}
{\cal H}_V = E_0 |v\rangle\langle v|,
\end{equation}
describing a quantum dot with energy of $E_0$. Here, $|v\rangle$ is
the ket-state corresponding to one of the discrete energy states
which are induced by the defects. It is assumed that this vacancy
state has been coupled with the continuum edge states.\par
The coupling interaction between the edge states and the defect
state is given by ${\cal H}_{EV}$:
\begin{equation}\label{Eq06}\begin{split}
{\cal H}_{EV} & = \int_{-\pi}^{+\pi} u_U (k_U) |k_U \rangle\langle v | dk_U\\
& + \int_{-\pi}^{+\pi} u_L (k_L)|k_L \rangle\langle v| dk_L + H.C.
\end{split}
\end{equation}
in which $H.C.$ stands for the Hermitian conjugate and the related
term ensures the hermiticity of the Hamiltonian. Also, $u_U(k_U)$
and $u_L(k_L)$ represent the coupling strengthes between the
discrete vacancy state of $|v\rangle$, and the continuum edge modes
of $|k_U\rangle$ and $|k_L\rangle$ at the upper and lower edges of
the considered ribbon, respectively. \par
The LS formulation of the formal scattering theory can be used to
calculate the transmission coefficient. This equation, which is a
key concept in quantum mechanics, particularly in the context of
scattering, reads
\begin{equation}\label{Eq07}
|\psi_{out}\rangle = |\psi_{in}\rangle + G_0(E) V|\psi_{out}\rangle,
\end{equation}
where $|\psi_{in}\rangle$ is the incoming or unperturbed wave state,
$|\psi_{out}\rangle$ is the full output or  scattered wave function,
$E$ is the energy of the specified scattering system,  $G_0(E)$ is
the free-particle Green's function, and $V$ is the interaction
potential.\par
By defining the transition operator, ${\cal T}(E)$, as
\begin{equation}\label{Eq08}
V|\psi_{out}\rangle = {\cal T}(E)|\psi_{in}\rangle,
\end{equation}
the LS equation can be rewritten as follows:
\begin{equation}\label{Eq09}
|\psi_{out}\rangle = |\psi_{in}\rangle + G_0(E) {\cal
T}(E)|\psi_{in}\rangle.
\end{equation}
By multiplying both sides of the above equation by $V$, it can be
converted into an equation to find the transition operator, ${\cal
T}(E)$, as:
\begin{equation}\label{Eq10}
{\cal T}(E) = V  + V G_0(E) {\cal T}(E).
\end{equation}
Usually, using the iteration method, the solution of the transition
operator can be written as an infinite expansion of the Green's
operator $G_0(E)$ and the interaction potential $V$ as
\begin{equation}\label{Eq11}
{\cal T}(E) = V + V G_0(E) V+ V G_0(E) V  G_0(E) V + \cdots.
\end{equation}
In the considered problem, the interaction potential $V$ is
equivalent to the term of ${\cal H}_{EV}$  in the entire Hamiltonian
and the Green operator, $G_0(E)$, is the associated to the rest of
the Hamiltonian, ${\cal H}_E + {\cal H}_V$. Consequently,  it is
possible to write:
\begin{equation}\label{Eq12}
G_0(E) = G_{E}(E) + G_V(E),
\end{equation}
in which
\begin{equation}\label{Eq13}\begin{split}
G_E(E) =  & \int_{-\pi}^{+\pi} {|k_U\rangle\langle k_U|\over E-
\hbar {\rm v}_F k_U+ i 0^+} dk_U\\ & \qquad\qquad+\int_{-\pi}^{+\pi}
{|k_L\rangle\langle k_L|\over E+ \hbar {\rm v}_F k_L+ i 0^+} dk_L,
\end{split}
\end{equation}
and
\begin{equation}\label{Eq14}
G_V = {|v \rangle\langle v|\over E-E_0+i0^+}.
\end{equation}
After some algebraic operation, it can be shown that
\begin{equation}\label{Eq15}
VG_0(E)VG_0(E)V = \alpha V,
\end{equation}
where $\alpha$ is given by the expression of
\begin{equation}\label{Eq16}\begin{split}
\alpha &= g_0(E) \Big(\int_{-\pi}^{+\pi} {\big|u_U(k_U)\big|^2 \over
E-\hbar {\rm v}_F k_U + i0^+} dk_U \\ &\qquad\qquad+
\int_{-\pi}^{+\pi} {\big|u_L(k_L)\big|^2 \over E+\hbar {\rm v}_F k_L
+ i0^+} dk_L\Big),
\end{split}
\end{equation}
in which
\begin{equation}\label{Eq17}
g_0(E) = {1 \over E- E_0+i0^+}.
\end{equation}
The combination of Eqs.~\eqref{Eq10} and~\eqref{Eq15} leads to
\begin{equation}\label{Eq18}
{\cal T}(E) = {V + V G_0(E) V \over 1- \alpha}.
\end{equation}
In the following calculation, it is assumed that the interaction
occurring during the scattering process is both short-range and
local. Also, for simplicity, it is assumed that $u_U(k_U)$ and
$u_L(k_L)$ are some constant functions denoted by $u_U$ and $u_L$,
respectively. If the interaction that leads to scattering is
completely local, this is an accurate assumption. But, since in the
specified system this interaction is not completely local, this
assumption is an approximation in our study. Of course, in the
following by comparing the simulation results with those obtained
from LB formulation, we will see that this approximation is very
good and efficient. Anyway, this assumption helps us to easily
calculate the integrals appearing in Eqs.~\eqref{Eq16}, so that we
have:
\begin{equation}\label{Eq19}\begin{split}
\int_{-\pi}^{+\pi} {1 \over E - \hbar {\rm v}_F k_U +i0^+} dk_U = {i\pi \over \hbar {\rm v}_F},\\
\int_{-\pi}^{+\pi} {1 \over E+\hbar {\rm v}_F k_L + i0^+} dk_L =
{i\pi \over \hbar {\rm v}_F}.
\end{split}
\end{equation}
With the above results, $\alpha$ is reduced to:
\begin{equation}\label{Eq20}
\alpha = \alpha_U + \alpha_L,
\end{equation}
where
\begin{equation}\label{Eq21}\begin{split}
\alpha_U & = g_0(E) \Big({\pi i \over \hbar{\rm v}_F}\Big) \big|
u_U\big|^2,\\ \alpha_L & = g_0(E) \Big({\pi i \over \hbar{\rm
v}_F}\Big) \big| u_L\big|^2.
\end{split}
\end{equation}
To obtain the transmission coefficient, an initial edge state is
considered which travels along the upper edge of the ribbon with an
initial momentum of $k_0$. Consequently, at a far distance before
the defect, the initial state can be represented as;
\begin{equation}\label{Eq22}
\langle \ell | \psi_{in} \rangle  = \langle \ell | k_0\rangle
=e^{ik_0\ell}.
\end{equation}
Here, the lattice constant is assumed as the unit of length and
$\ell \to -\infty$. To know the transmission probability, we should
calculate the probability of detecting the particle at a far
distance after the central scattering region and along the upper
side of the ribbon. If the particle is detected in this distance, it
means that the output wave function can be displayed as
\begin{equation}\label{Eq23}
\langle \ell | \psi_{out} \rangle  = t \langle \ell | k_0\rangle = t
e^{ik_0\ell},
\end{equation}
in which $t$ is the transmission amplitude and $\ell\to +\infty$.
Multiplying both sides of~Eq.\eqref{Eq09} by bra-basis of
$\langle\ell|$ reads:
\begin{equation}\label{Eq24}
\langle\ell|\psi_{out}\rangle = \langle\ell |\psi_{in}\rangle +
\langle\ell | G_0(E) {\cal T}(E)|\psi_{in}\rangle.
\end{equation}
Using the fact that for the considered case
\begin{equation}\label{Eq25}
V = \int_{-\pi}^{+\pi} u_U(k_U) |k_U\rangle \langle v | d k_U +
H.C.,
\end{equation}
and
\begin{equation}\label{Eq26}
G_0(E) = \int_{-\pi}^{+\pi} { |k_U\rangle \langle k_U | \over
E-E_0+i0^+} d k_U,
\end{equation}
it can be shown that
\begin{equation}\label{Eq27}\begin{split}
\langle\ell | G_0(E) {\cal T}(E)|k_0\rangle = &{ g_0(E)\over 1-\alpha}\\
\times \int_{-\pi}^{+\pi} & {u_U^*(k_U)u_U(k_0)e^{ik_0\ell}  \over
E-i{\rm v}_F k_U+i0^+} d k_U.
\end{split}
\end{equation}
Considering that $u_U(k_U)$ and $u_U(k_0)$ can be assumed equal to a
constant value of $u_U$, the following closed form expression can be
derived
\begin{equation}\label{Eq28}
\langle\ell | G_0(E) {\cal T}(E)|k_0\rangle = {2\alpha_U\over
1-\alpha} e^{ik_0\ell},
\end{equation}
Finally, by inserting Eqs.~\eqref{Eq22},~\eqref{Eq23},
and~\eqref{Eq28} in Eq.~\eqref{Eq24}, and performing some algebraic
calculation, the transmission amplitude is drivable as
\begin{equation}\label{Eq29}
t = 1+ {2 \alpha_U \over 1-\alpha}={1+\alpha_U-\alpha_L \over
1-\alpha_U-\alpha_L}.
\end{equation}
If the coupling between the discrete vacancy and the continuous
upper-edge states is absent, we have $u_U = 0$ and consequently
$\alpha_U=0$.  Then, as expected and is clear from the above
equation, $t=1$ which result in a perfect transmission. On the other
hand,  if there is no coupling between the discrete vacancy and the
continuous lower-edge states, then $u_L=0$ and $t$ is reduces to
\begin{equation}\label{Eq30}
t = {1 + \alpha_U \over 1-\alpha_U}.
\end{equation}
Since $\alpha_U$ is purely imaginary,  $T(E) = |t|^2=1$ and the
transmission is complete, as is expected.  But in the interaction
with the lower edge, there is a possibility of reflection and
scattering to the lower wire. This discussion confirms that for
observing the topological Fano resonance in a Kane-Mele nanoribbon,
it is essential that the vacancy discrete states becomes coupled
with the continuous edge modes on both upper and lower sides. Also,
the above fact indicates that the strength of coupling between
discrete and continuous modes affects the quality of observing Fano
resonance.\par
\begin{figure}[t!]
\begin{center}
\includegraphics[scale=0.72]{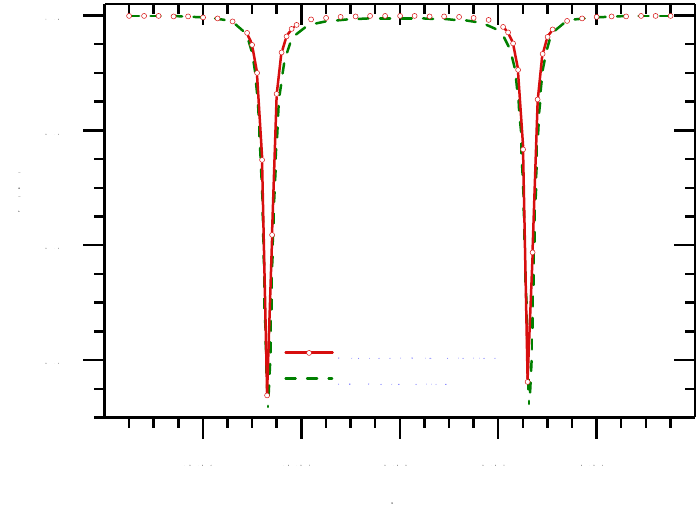}
\end{center}
\caption{ Comparison of the transmission spectrum of the specified
system obtained from the current simulation based on the
LS~scattering formalism and the results obtained from LB
formulation. This comparison verifies the efficiency and accuracy of
the simulation method presented in this study confirming the
occurence of the topological Fano resonance in the considered
structur.~\label{Fig04}}
\end{figure}
In Fig.~\ref{Fig04}, the analytically simulated transmission
probability, $T(E)$, is plotted as a function of energy, $E$, and is
compared with the calculations obtained from LB formalism. As
expected, the two graphs are both qualitatively and quantitatively
very similar and this figure correctly shows the behavior exhibited
in Fig.~\ref{Fig02}. The very good match of the two graphs shows
that the assumption that $u_U(k_U)$ and $u_L(k_L)$ are constant is
an approximation that is very close to being accurate. This figure
shows that despite the simplicity of the carried simulation based on
LS~scattering formulation, it is reasonably efficient and accurate
in quantitative and qualitative prediction of the topological Fano
resonance in Kane-Mele nanoribbons.\par
\section{Summary and Conclusion~\label{Sec04}}
A nanoribbon of a two-dimensional hexagonal lattice with finite
width and armchair edges, which governed by the Kane-Mele
Hamiltonian, was considered. This Hamiltonian describes the
spin-orbit interaction which leads to the quantum spin Hall effect.
It was assumed that the length of the ribbon at the far distances is
limited to two left and right leads. As a result, at the armchair
sides, edge continuum states, which are topologically protected,
flowed to the left and right with Fermi velocities. It was assumed
that the length of the ribbon, at long distances, was limited to two
left and right leads. As a result, at the armchair sides, edge
continuum states, which are topologically protected, traveled to the
left and right with Fermi velocities. This situation was similar to
considering the ribbon's sides as two current-carrying wires. By
creating a regular array of on-site vacancies on a zigzag chain
across the width of the ribbon, a conduction channel between the
edge wires of the ribbon was created. This cannel caused the
creation of a central scattering region around the zigzag chain that
contained the vacancy defects. Also, this conduction channel, which
is due to the overlap of zero-energy localized states surrounding
the vacant sites, allows the creation of discrete vacancy states,
the coupling of these discrete states with continuous edge states,
and the occurrence of backscattering events. It saw shown that the
coupling of topologically protected continuous edge states and the
discrete vacancy states and their interference with each other made
it possible to observe topological Fano resonance. First, the Fano
topological resonance profile was produced using the LB formulation.
Then, employing the LS~scattering formalism and using the Green's
operator as well as the transition matrix  this profile was
simulated, analytically. In the calculations, it was shown that the
transmission spectrum of the desired system is a function of the
energy of the system and the degree of coupling between the discrete
vacant states with the continuous edge states on both sides of the
ribbon. By adjusting the width of the ribbon, choosing the number of
vacancies, and checking how they are arranged on the zigzag chain
across the ribbon, it is possible to adjust and control the
vacancy-induced topological Fano resonance in the Kane-Mele
nanoribbons. From an analytical perspective, the simulation of
topological Fano resonance involves using the LS scattering
framework. This research provides a benefit by performing associated
calculations analytically, producing outcomes that closely match
those obtained using the LB approach, showcasing both quantitative
and qualitative precision. These findings contribute to the
understanding of the impact of vacancies on the electronic and
transport properties of nanoribbons, which is essential for the
design, control, and sensing applications of such systems.
\begin{acknowledgments}
The first author, SJ, would like to acknowledge the office of
graduate studies at the University of Isfahan for their support and
research facilities. Additionally, the fourth author, MA, would like
to acknowledge the support received from the Abdus Salam (ICTP)
associateship program.
\end{acknowledgments}

\end{document}